# New results from RENO and prospects with RENO-50

Soo-Bong Kim

*KNRC, Department of Physics and Astronomy, Seoul National University, Seoul 151-742, South Korea*

**Abstract**

RENO (Reactor Experiment for Neutrino Oscillation) has made a definitive measurement of the smallest mixing angle $\theta_{13}$ in 2012, based on the disappearance of electron antineutrinos. More precise measurements have been obtained and presented on the mixing angle and the reactor neutrino spectrum, using ~800 days of data. We have observed an excess of IBD prompt spectrum near 5 MeV with respect to the most commonly used models. The excess is found to be consistent with coming from reactors. We have also successfully measured the reactor neutrino flux with a delayed signal of neutron capture on hydrogen. A future reactor neutrino experiment of RENO-50 is proposed to determine the neutrino mass hierarchy and to make highly precise measurements of $\theta_{12}$, $\Delta m_{21}^2$, and $\Delta m_{31}^2$. Physics goals and sensitivities of RENO-50 are presented with a strategy of obtaining an extremely good energy resolution toward the neutrino mass hierarchy.

## 1. Oscillation of reactor antineutrinos

A fission reactor is a copious source of electron antineutrinos produced in the $\beta$ decays of neutron-rich nuclei. Nuclear reactors have played crucial roles in experimental neutrino physics. The discovery of the neutrinos was made at the Savannah River reactor in 1956 [1]. The KamLAND collaboration observed disappearance of reactor neutrinos and distortion in the energy spectrum due to neutrino oscillations [2]. Daya Bay, Double-Chooz, and RENO collaborations determined the smallest mixing angle $\theta_{13}$ based on the observed disappearance of reactor neutrinos [3, 4, 5]

In the present framework of three flavors, neutrino oscillation is described by a unitary Pontecorvo-Maki-Nakagawa-Sakata matrix with three mixing angles ($\theta_{12}$, $\theta_{23}$ and $\theta_{13}$) and one CP phase angle [6, 7]. Neutrino oscillation was discovered in the atmospheric neutrino by the Super-Kamiokande experiment in 1998 [8]. The next round of neutrino experiments are under consideration or preparation to determine the CP violation phase and the neutrino mass hierarchy.

A few MeV, low-energy reactor neutrinos have relatively short oscillation lengths to compensate for rapid reduction of the neutrino flux at a distance. Reactor neutrino measurements can determine the mixing angle without the ambiguities associated matter effects and CP violation. The reactor neutrino detector is not necessarily large, and construction of a neutrino beam is not needed. Past reactor experiments had a single detector located about 1 km from reactors. The new generation reactor experiments, Daya Bay and RENO, have significantly reduced uncertainties associated with the measurement of $\theta_{13}$ using two identically performing detectors at near and far locations from reactors. The exciting results provide a comprehensive picture of neutrino transformation among three kinds of neutrinos. An accurate value of $\theta_{13}$ by the reactor experiment will be able to offer the first glimpse of the CP phase angle, if combined with a result from an accelerator neutrino beam experiment [9].

## 2. The RENO experiment

RENO was the first reactor experiment to take data with two identical near and far detectors in operation, from August 2011. In early April 2012, the experiment successfully reported a definitive measurement of $\theta_{13}$ by observing the disappearance of reactor neutrinos [5].

RENO detects antineutrinos from six reactors at Hanbit Nuclear Power Plant in Korea. The six pressurized water reactors with each maximum thermal output of 2.815 GW$_{th}$ (reactors 3, 4, 5 and 6) or 2.775 GW$_{th}$ (reactors 1 and 2) are lined up in roughly equal distances and in a span of ~1.3 km. The identical near and far antineutrino detectors, each having 16 tons of Gadolinium (Gd) loaded liquid scintillator (LS) as a neutrino target, are located at 294 m and 1383 m, respectively, from the center of the reactor arrays. The far (near) detector is under a 450 (120) meters of water equivalent rock overburden. The reactor-flux weighted baseline is 408.56 m for the near detector, and 1443.99 m for the far detector.

A nuclear reactor produces about $10^{20}$ antineutrinos per GW and per second, mainly coming from the beta decays of fission products of $^{235}$U, $^{238}$U, $^{239}$Pu, and $^{241}$Pu. The reactor antineutrino is detected via the inverse beta decay (IBD) reaction, $\nu_e + p \rightarrow e^+ + n$. The coincidence of a prompt positron signal and a delayed signal from neutron capture by Gd provides the distinctive IBD signature.

The RENO detector consists of a main inner part (ID) and an outer veto part (OD). The ID is contained in a cylindrical stainless steel vessel that houses two nested cylindrical acrylic vessels. The inner most acrylic vessel holds 16 tons of ~0.1% Gd-doped LS. It is surrounded by a gamma-catcher region with Gd-unloaded LS inside an outer acrylic vessel. Outside the gamma catcher is a buffer region with mineral oil. Light signals emitted from particles interacting in the ID are detected by 354 10-inch photomultiplier tubes (PMTs), providing 14% surface coverage, that are mounted on the inner wall of the stainless steel container. The OD consists of highly purified 1.5 m thick water layer in order to identify events coming from outside by their Cherenkov radiation and to shield against ambient γ-rays and neutrons from the surrounding rocks. The OD is equipped with 67 10-inch water-proof PMTs mounted on the wall of the veto vessel.

The energy calibration is performed with γ-rays coming from radioactive sources of $^{137}$Cs (E$_\gamma$=0.622 MeV), $^{68}$Ge (E$_\gamma$=1.022 MeV), $^{60}$Co (E$_\gamma$=2.506 MeV), and Am-Be (E$_\gamma$=4.945 MeV due to neutron capture on C), and from IBD delayed signals of neutron capture on H (E$_\gamma$=2.223 MeV) or Gd (E$_\gamma$=7.937 MeV). A non-linear response of the scintillating energy is obtained from the several calibration samples and fairly well explained by a commonly used modelling function of $E_{vis}/E_{true}=a+b/[1+exp(-c \cdot E_{true}+d)]$ as shown in Fig. 1. The energy scale uncertainty is better than 0.5% in IBD prompt energies. Agreement of energy scale between the near and far detectors is found to be better than 0.1%.

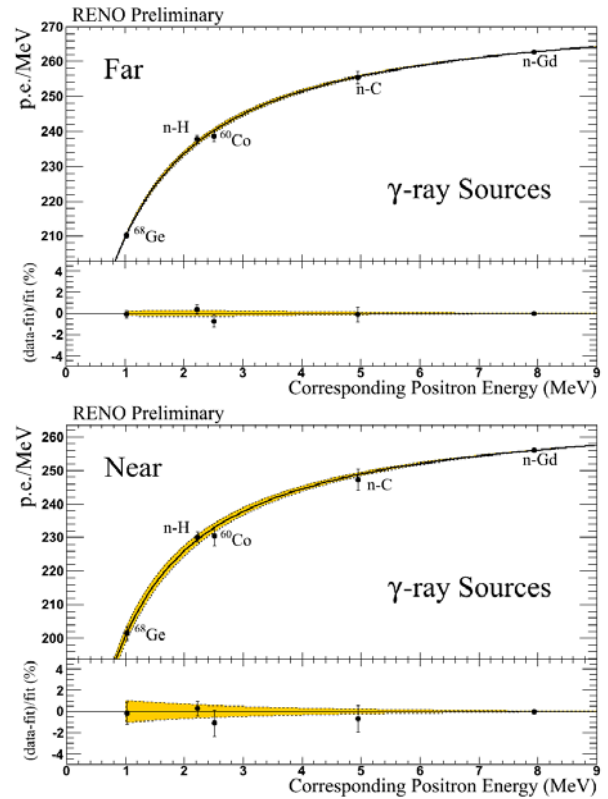

Fig. 1. A non-linear response of scintillating energy. A commonly used modeling is applied to describe the visible energies of γ-rays coming from several radioactive sources and delayed signals. The energy scale uncertainty is better than 0.5% in IBD prompt energies.

The visible energy of a positron is determined by the observed photoelectrons corresponding to 1.02 MeV of two γ-rays from the positron annihilation plus the positron kinetic energy. Figure 2 shows excellent agreement of MC and data in the electron kinetic energy of β-decays from unstable isotopes of $^{12}$B and $^{12}$N that are produced by cosmic muons.

RENO has collected more than 1,100 live days of data as of November 2014, to observe ~1.5M reactor neutrino events in the near detector and ~0.15M events in the far detector. In this workshop, we report the $\theta_{13}$ measurement results from roughly 800 live days of data sample taken through the end of 2013.

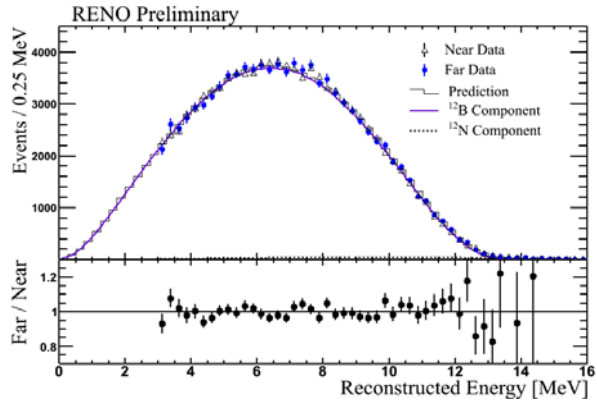

Fig. 2. Comparison of MC and data in the electron kinetic energy of β-decays from unstable isotopes of $^{12}$B and $^{12}$N. Excellent agreement is seen between near detector and far detector.

## 3. Results on the neutrino mixing angle $\theta_{13}$

Two identical antineutrino detectors are located at near and far locations from reactors to allow a relative measurement from their measured neutrino rates. The measured far-to-near ratio of antineutrino fluxes can considerably reduce systematic errors coming from uncertainties in the reactor neutrino flux, target mass, and detection efficiency.

There are three remaining backgrounds in the final IBD candidate sample: (1) uncorrelated background of accidental coincidence from random association of prompt- and delayed-like events due to external gamma rays from radioactivity in the surrounding rock and detector noise events (accidentals), (2) correlated background due to energetic neutrons that are produced by cosmic muons traversing the surrounding rock and the detector, enter the ID, and interact in the target to produce a recoil proton as a prompt-like signal (fast neutrons), and (3) correlated background due to unstable isotopes of $^9$Li/$^8$He that are produced by cosmic muons and decay into a positron and a neutron ($^9$Li/$^8$He β-n emitters).

### 3.1. Measured $\theta_{13}$ with neutron captures on Gd

After applying all selection criteria we have obtained a total of 457,176 (53,632) IBD candidate events with the prompt energy less than 8 MeV, using 761 (795) live days of data in the near (far) detector. In the final IBD candidate sample, the remained background is estimated to be 8.1% (3.1%) of the total far (near) events. The average daily observed IBD rates after subtracting backgrounds are 582.06±1.37 /day and 61.99±0.45 /day for the near and the far detectors, respectively. Fig. 3 presents the measured mean daily-rates of IBD candidates after background subtraction in the near and far detectors as a function of the expected rate for different reactor power conditions. The expected rates are obtained from the oscillation effect of the best fit, and well agree with the measured rates. The extrapolation to the zero reactor power of the fit to data demonstrates an excellent background estimation and subtraction.

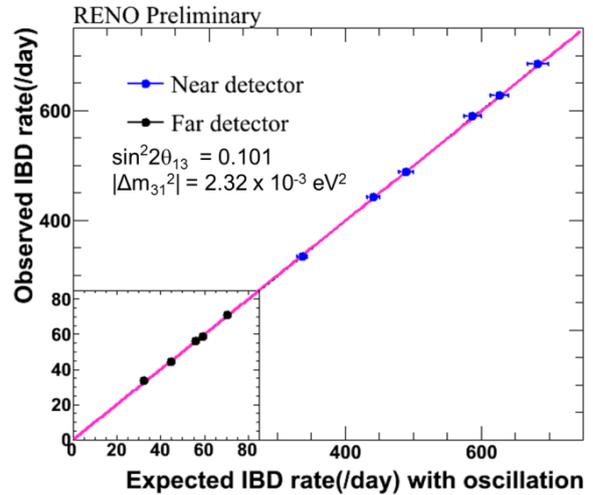

Fig. 3. Measured daily number of reactor neutrinos as a function of the expected number that is obtained for the grouped data samples with similar reactor power conditions. The solid lines are the best fits to the data, demonstrating an excellent background subtraction.

The RENO collaboration reported a definitive measurement of $\theta_{13}$ based on 220 days of data taken through March 2012 in April 2012 [5], and has presented updated results in several conferences later. Improved values of $\theta_{13}$ were reported based on ~400 live days of data through October 2012 at Neutrino Telescope 2013 workshop [10] and at TAUP 2013 conference [11]. A more precise measurement was reported in the Neutrino 2014 conference and presented in this workshop, as $\sin^2(2\theta_{13}) = 0.101 \pm 0.008(\text{stat.}) \pm 0.010(\text{syst.})$, based on ~800 live days of data through December 2013 [12].

The history of $\theta_{13}$ measurements by RENO is shown in Table 1. The improvements came from better understanding of the detector energy scale, more accurate estimation of a cosmic-ray induced background uncertainty, and more data. Additional efforts have been made to reduce the systematic errors, and more precise measurement result is expected to be released in the near future.

Table 1

Summary of measured values of $\theta_{13}$ by the RENO experiment

| Measured values of $\sin^2(2\theta_{13})$ | Data sample (Aug. 2011~ ) | Refs. (year) |
|---|---|---|
| 0.113±0.013(stat.)±0.019(syst.) | ~220 days (~ Mar. 2012) | PRL 108 (2012) [5] |
| 0.100±0.010(stat.)±0.015(syst.) | ~400 days (~ Oct. 2012) | NuTel (2013) [10] |
| 0.100±0.010(stat.)±0.012(syst.) | ~400 days (~ Oct. 2012) | TAUP (2013) [11] |
| 0.101±0.008(stat.)±0.010(syst.) | ~800 days (~ Dec. 2013) | Neutrino (2014) [12] |

### 3.2. Measured $\theta_{13}$ with neutron captures on H

RENO has also measured the value of $\theta_{13}$ from the IBD events that are associated with a delayed signal of a neutron capture on hydrogen (n-H). This is possible due to reasonably low accidental backgrounds as a result of successful radioactivity-reduction in the LS and PMT.

The first measurement of $\theta_{13}$ using a ~400 day n-H data sample was reported in the Neutrino 2014 conference as $\sin^2(2\theta_{13})$ = 0.095 ± 0.015(stat.) ± 0.025(syst.) [12]. An improved result is presented in this workshop as $\sin^2(2\theta_{13})$ = 0.103 ± 0.014(stat.) ± 0.014(syst.). The significantly reduced systematic error comes from complete removal of multiple neutron backgrounds and more precise estimation of accidental backgrounds.

We have been continuing efforts to reduce the systematic errors in this measurement. A more precise measurement of $\theta_{13}$ is soon expected to be comparable to the result with a delayed signal of neutron capture on Gd. Combining those results of two independent measurements, we may obtain quite accurate values of the mixing angle $\theta_{13}$ and the squared mass difference $\Delta m_{ee}^2$.

### 4. 5 MeV excess in the reactor neutrino spectrum

The RENO near and far detectors have neutrino flight distances of ~300m to ~1.5 km depending on reactors, and can determine the squared mass difference $|\Delta m_{ee}^2|$ based on the survival probability of electron antineutrinos. The total background rate is estimated to be 17.96±1.00 (near) or 4.61±0.31 (far) events per day in the 800 day sample. The expected rate and spectrum of reactor antineutrinos are calculated for duration of physics data-taking, taking into account the varying thermal powers and fission fractions of each reactor.

RENO has obtained an unprecedentedly accurate measurement of the reactor neutrino flux and spectrum. Fig. 4 shows the observed spectra of IBD prompt signals for the near and far detectors after subtracting backgrounds, compared to the MC expectations from the best fit parameters to neutrino oscillation. A clear spectral difference from the current reactor neutrino models [14, 15], is observed at 5 MeV with excess magnitudes, 2.2±0.1(stat.)±0.4(syst.)% of the total observed reactor neutrino flux in the near detector or 1.8±0.3(stat.)±0.6(syst.)% in the far detector. The systematic error was estimated by uncertainties of energy scale, normalization, isotope fraction, MC modeling, and oscillation parameters. Including the expected spectral shape error of 0.5% from the reactor models, the significance of the shape difference is more than 3.5σ. We observe for the first time that the

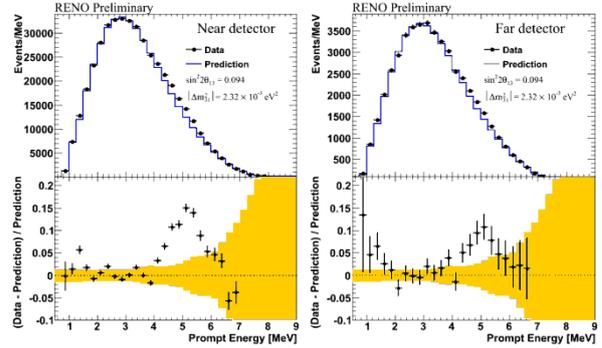

Fig. 4. Comparison of observed and expected IBD prompt energy spectra. A shape difference is cleary seen at 5 MeV. The observed excess is correlated with the reactor power, and corresponds to 2.2% of the total observed reactor neutrino flux.

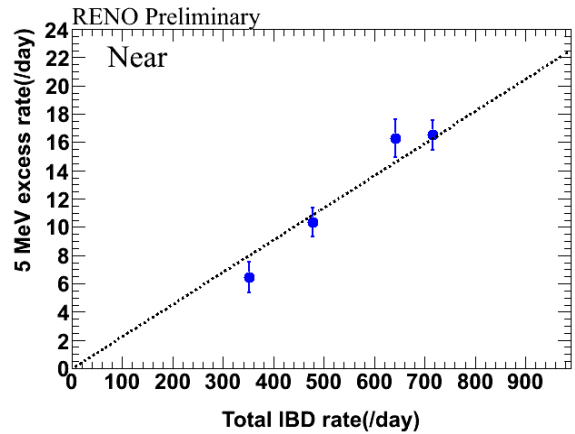

Fig. 5. Correlation between the 5 MeV excess daily rate and the expected IBD daily rate with oscillations in the near detector. This indicates the excess is strongly proportional to the thermal power.

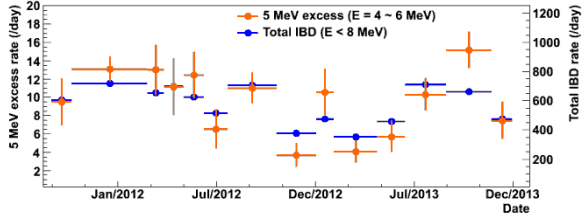

Fig. 6. Correlation between the 5 MeV-excess daily rate and the total observed IBD daily rate in the near detector, for each data-taking time period..

5 MeV excess is strongly proportional to the reactor power as shown in Fig. 5, indicating that the excess of IBD events comes from the reactors.

Fig. 6 show a clear correlation between the 5 MeV excess daily rate and the total observed IBD daily rate in the near detector for each data-taking period. This again indicates the 5 MeV excess indeed comes from the reactors.

## 5. Implications and future prospects of precise measurement of $\theta_{13}$

RENO has definitively measured the value of $\theta_{13}$ by the disappearance of electron antineutrinos. Based on unprecedentedly copious data, RENO has performed rather precise measurements of the value. The exciting result provides a comprehensive picture of neutrino transformation among three kinds of neutrinos and opens the possibility of search for CP violation in the leptonic sector. The surprisingly large value of $\theta_{13}$ will strongly promote the next round of neutrino experiments to find CP violation effects and determine the neutrino mass hierarchy. The successful measurement of $\theta_{13}$ has made the very first step on the long journey to the complete understanding of the fundamental nature and implications of neutrino masses and mixing parameters.

The total systematic errors of RENO, $\delta\sin^2(2\theta_{13})=\pm 0.010$, mainly come from three uncertainties of reactor neutrino flux, detector efficiency and backgrounds. The uncorrelated uncertainty of reactor flux among reactors is 0.9% (0.7%: fission fraction, 0.5%: thermal power), and results in $\delta\sin^2(2\theta_{13}) = \pm 0.0032$. The uncorrelated uncertainty of detection efficiency between the near and far detectors is 0.2%, and contributes additional $\delta\sin^2(2\theta_{13}) = \pm 0.0032$. The background uncertainty is estimated as 6.7% (5.6%) mainly due to $^{252}$Cf source contamination (cosmic $^9$Li/$^8$He background) in the far (near) detector, and dominates the total systematic error by $\delta\sin^2(2\theta_{13}) = \pm 0.0089$. The background estimation is entirely based on the control data samples, and thus the uncertainty is expected to be reduced with more data.

Precise measurements of $\theta_{13}$ by the reactor experiments will provide the first glimpse of the CP phase angle if accelerator beam results are combined. Based on total 5 years of data, the RENO experiment is expected to obtain a measured $\sin^2(2\theta_{13})$ value with a precision of 7% according to the design goal. With a better understating of systematic uncertainties, it could become as good as 5%, and can be even smaller if the n-H result is going to be combined. We will also make a direct measurement of $|\Delta m_{ee}^2|$ from the energy dependent oscillation effects in the near future, and will make a more precise measurement of reactor neutrino spectrum.

## 6. RENO-50: future reactor experiment for neutrino mass hierarchy

An underground detector of RENO-50 [13] under proposal will consist of 18,000 tons of ultra-low-radioactivity liquid scintillator and 15,000 high quantum efficiency 20" photomultiplier tubes, located at roughly 50 km away from the Hanbit nuclear power plant in Korea where the neutrino oscillation due to $\theta_{12}$ takes place at maximum (see Fig. 7).

The detector is expected to detect neutrinos from nuclear reactors, the Sun, Supernova, the Earth, any possible stellar object and a J-PARC neutrino beam as well. It will be a multi-purpose and long-term operational detector, and also a neutrino telescope. The main goal is to determine the neutrino mass hierarchy and to measure the unprecedentedly accurate (<1%) values of $\theta_{12}$, $\Delta m_{21}^2$, and $\Delta m_{31}^2$. It is also expected to detect ~5,600 events of a neutrino burst from a Supernova in our Galaxy, ~1,000 geo-neutrino events for 6 years, and ~200 events of muon neutrinos from the J-PARC beam every year.

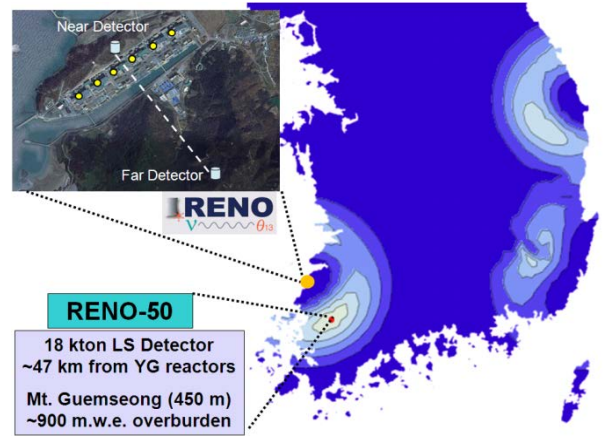

Fig. 7. The RENO-50 detector will be located at underground of Mt. Guemseong in a city of Naju, 47 km from the Hanbit nuclear power plant. The contours of different colors indicate the sensitivity of mass hierarchy determination. The perpendicular direction from the reactor alignment has the highest sensitivity.

The RENO-50 would be able to observe the manifestation of mass hierarchy in the oscillation effect if it establishes an extremely good energy resolution of ~3% at 1 MeV. The energy resolution could be achieved based on maximized light yield of LS larger than 1,000 photoelectrons per MeV, through making (a) an increased photosensitive area using 15,000 20" PMTs, (b) use of high (35%) quantum efficiency PMTs, and (c) an increased attenuation length of LS up to 25m.

The high precision measurements of $\theta_{12}$, $\Delta m_{21}^2$, and $\Delta m_{31}^2$ can make a strong impact on explaining the pattern of neutrino mixing and its origin. It will also provide useful information on the effort of finding a flavor symmetry. A RENO-50 proposal has been submitted for full construction funding. A R&D funding is allocated from the end of 2014, and will continue in the next 3 years. R&D efforts will be made on demonstrating feasibility of 3% energy resolution at 1 MeV, essential for determining the neutrino mass hierarchy. If the construction funding is timely made, we expect to start the experiment in 2021.

## Acknowledgments

This work is supported by the Ministry of Science, ICT and Future Planning of Korea and the Korea Neutrino Research Center selected as a Science Research Center by the National Research Foundation of Korea (NRF).